# Spectroscopic evidence of disorder-induced quantum phase transitions in monolayer Fe(Te,Se) superconductor

Guanyang He,[1,2] Ziqiao Wang,[1,3] Longxin Pan,[4] Yuxuan Lei,[1] Fa Wang,[1] Yi Liu,[4,9] Nandini Trivedi,[5] and Jian Wang[1,3,6,7,8*]

[1]*International Center for Quantum Materials, School of Physics, Peking University, Beijing 100871, China*

[2]*ShanghaiTech Laboratory for Topological Physics, ShanghaiTech University, 201210 Shanghai, China*

[3]*Quantum Science Center of Guangdong-Hong Kong-Macao Greater Bay Area (Guangdong), Shenzhen 518045, China.*

[4]*School of Physics and Beijing Key Laboratory of Opto-electronic Functional Materials & Micro-nano Devices, Renmin University of China, Beijing 100872, China.*

[5]*Department of Physics, The Ohio State University, Columbus, OH 43210*

[6]*Beijing Key Laboratory of Quantum Devices, Peking University, Beijing 100871, China*

[7]*Hefei National Laboratory, Hefei 230088, China*

[8]*Collaborative Innovation Center of Quantum Matter, Beijing 100871, China*

[9]*Key Laboratory of Quantum State Construction and Manipulation (Ministry of Education), Renmin University of China, Beijing 100872, China.*

*Corresponding author.

Jian Wang (jianwangphysics@pku.edu.cn)

## Abstract

The superconductor-insulator transition as a paradigm of quantum phase transitions has attracted tremendous interest over the past three decades. While the magnetic field and carrier density can be tuned to drive the transition, the role of disorder in the transition is not well understood due to the complicated interplay between superconductivity and electron localization. In this work, we controllably introduce disorder in a two-dimensional high-temperature superconductor by depositing iron clusters onto the superconducting monolayer Fe(Te,Se) crystalline film. The spectral evolution from superconducting gaps to insulating gaps with increasing disorder is detected by scanning tunneling spectroscopy measurements. When the disorder is strong, large U-shaped gaps are observed and attributed to the localization-enhanced Cooper pair correlation. Our observations provide the insight into the emergent phases of low-dimensional and high-temperature superconductors with disorder.

A quantum phase transition describes the transition between different ground states at zero temperature, which can be accessed by tuning a parameter such as the magnetic field, carrier density, or disorder [1-4]. Upon increasing disorder, it is possible to drive a superconductor to insulator transition (SIT) in superconducting thin films. It was found theoretically [5-8] that the superconducting order parameter becomes increasingly inhomogeneous with disorder on the scale of the coherence length, forming superconducting puddles. The local density of states (LDOS) in the puddles shows superconducting gaps with coherence peaks, whereas the LDOS outside the puddles may exhibit larger spectral gaps without coherence peaks due to incoherent Cooper pairs with large phase fluctuations. A persisting spectral gap with suppressed coherence peaks has been observed as disorder increases in conventional superconductors [9, 10], and subsequent theoretical [11, 12] and experimental studies [11, 13] further revealed the spatial inhomogeneity of spectral gap and its connection with suppressed superfluid stiffness. In this scenario, SIT is driven by the vanishing of superfluid stiffness as a consequence of suppressed Josephson coupling between superconducting puddles [7, 14-16], during which the single-particle gap remains finite in space. The theory further predicted that the single-particle gap shows a non-monotonic behavior with disorder, decreasing initially but then growing with disorder at higher disorder near SIT and beyond [6, 7]. The multifractal nature of electron wavefunctions [17, 18] near SIT could modify the exponents that describe the growth of the gap with disorder [17, 19]. This unusual gap enhancement is suggested to originate from the enhanced electron pairing within the localization length. Since the localization length decreases with increasing disorder, it results effectively in enhanced electron pairing. Remarkably, it is theoretically pointed out that the spectral evolution across the SIT is distinct from both paradigms of Anderson localization that yields gapless localized states at the chemical potential, and of BCS superconductivity where the superconducting state is destroyed by gap closing.

Experimentally, comparing to transport measurements [1, 4], scanning tunneling microscopy/spectroscopy (STM/S) studies to understand SIT at the microscopic level are relatively rare. Several previous spectroscopic studies have investigated the disorder effect on conventional superconducting films [9, 10, 13, 20, 21], focusing on the superconducting regime or the critical regime near SIT. Instead, our experiments focus on the SIT extending to the insulating regime in the two-dimensional (2D) single-crystal iron-based monolayer showing high-transition temperature ($T_C$) superconductivity. In Fe(Te,Se) (FTS) superconducting monolayer, we reveal the evolution of spectral gaps in both the superconducting and insulating regimes of SIT. Superconducting FeSe films possess a rich phase diagram where superconductivity is highly tunable [22-26]. Through the substitution of Se by Te, monolayer FTS on $SrTiO_3$ (STO) is synthesized and has become a candidate to realize high-$T_C$ topological superconductor [27-29]. Moreover, relatively strong fluctuations due to the reduced dimensionality as well as the fine tunability of superconducting properties make monolayer FTS an ideal platform to study SITs. We grow superconducting FTS crystalline monolayer on STO substrates by molecular beam epitaxy (MBE), then introduce disorder controllably by *in situ* depositing iron clusters onto the film. The superconducting properties and disorder-induced spectroscopic changes in monolayer FTS are systematically examined by STM/S at 4.3 K (see experimental details in Supporting Information), well below the $T_C$ of monolayer FTS. Significantly, our results show that with the increasing amounts of disorder, the superconducting gap evolves to a V-shaped gap, then to a large U-shaped spectral gap, demonstrating the intriguing interplay between Anderson localization and superconductivity.

Figure 1(a) shows the STM topography of pristine monolayer FTS grown on STO terraces. The thickness of monolayer FTS is around 0.59 nm (see Fig. S1 in Supporting Information), indicating a Te concentration around 50% [30]. Figure 1(b) shows the normalized spectra $G_N$ of monolayer FTS at varying temperatures (see Supporting Information for normalization methods), where $G_N$ exhibits a well-identified superconducting gap and coherence peaks at 4.3 K. The gap-filling temperature $T^*$ is estimated by extrapolating the zero-bias conductance (ZBC) of $G_N$ in Fig. 1(b) to unity with a linear fitting, giving $T^* \approx 56.4$ K as shown in Fig. 1(c), which is comparable to the reported $T_C$ of monolayer FTS/STO [28, 30]. After depositing iron clusters onto the sample by MBE, the clusters scatter over the film surface randomly as bright spots. The typical topography is shown in Fig. 1(d). All bright spots with heights of 60 pm or higher above the FTS surface are considered clusters (see cluster topographies in Fig. S3 in Supporting Information), since the height of an iron adatom on monolayer FeSe is around 62 pm [31]. On this basis, the surface cluster coverage of Fig. 1(d) (number of clusters per area, see the counting process in Fig S4 of Supporting Information) is estimated around 0.027 $nm^{-2}$. This cluster coverage is merely an indicator to exhibit the gradual increment of clusters in our consecutive deposition of iron clusters. After iron clusters are deposited for the first time, such local coverage varies on a macroscopic scale (millimeter) on the sample, within a range of 0.013 ~ 0.023 $nm^{-2}$ (see Fig. S5(a1)-(a5) in Supporting Information) after iron clusters are deposited for the 1st time. The range of coverage is raised to 0.013 ~ 0.038 $nm^{-2}$ (see Fig. S5(b1)-(b5) in Supporting Information) after iron clusters are deposited for the 2nd time. Figure 1(e) exhibits the spectra for monolayer FTS with various cluster coverages, which are taken on an FTS area at least 5 nanometers away from any clusters to reveal the influence of disorder on the superconducting properties of monolayer FTS itself. Based on our observations, the spectral features in the field of view of a topographic image (around 40×40 $nm^2$) are consistent, and largely related to the cluster coverage. For a low disorder level (coverage around 0.017 $nm^{-2}$) the spectra of monolayer FTS show reduced superconducting gap and weaker coherence peaks. For a moderate disorder level (coverage around 0.027 $nm^{-2}$), the spectra become V-shaped with in-gap conductance and no clear coherence peaks.

Figure 1(f)-(g) shows the normalized spectra $G_N$ at varying temperatures for the region with the cluster coverage of 0.017 $nm^{-2}$ and 0.027 $nm^{-2}$, respectively. The corresponding values of $T^*$ are present in Fig. 1(h), obtained by the same extrapolating method of Fig. 1(c). At the cluster coverage of 0.017 $nm^{-2}$, $T^*$ = 46.7 K lower than that of the pristine monolayer (56.4 K) is pertinent to the suppressed superconductivity. Nevertheless, for the cluster coverage of 0.027 $nm^{-2}$, $T^*$ = 60.1 K is unexpectedly larger than that in the less-disordered situations. The fact that $T^*$ does not decrease monotonously with increasing disorders might indicate the appearance of pseudo-gaps at moderate cluster coverage, which are typically seen in amorphous superconductors near SIT [10, 20, 32]. In monolayer FeSe/STO, a lower zero-resistance $T_C$ compared to $T^*$ is observed because of the strong superconducting fluctuations in a 2D high-$T_C$ superconductor, where long-range phase coherence forms at a lower temperature than the temperature where local pairing starts to form [33, 34]. By raising the disorder level, we observe an enlarged T* at moderate disorder level, suggestive of further enhanced local pairing.

The dI/dV mappings are measured in an area with an accumulation of iron clusters, as shown in Fig. 2(a). For the mapping at the energy of -40 meV below the Fermi level in Fig. 2(b), patches of high LDOS (red color) show up especially on and near the clusters, amid a sea of depleted LDOS (blue color). Such a spatially inhomogeneous LDOS is indicative of electrons with energies below the mobility edge, where the electron scattering in a disordered medium leads to Anderson localization. To exclude the LDOS of iron cluster itself and study that of monolayer FTS in the dI/dV mappings, we remove the dI/dV values on the positions of clusters (dashed circles in Fig. 2(a)), as indicated by the black holes in Fig. 2(c)-(d). For the mappings at higher energies above the mobility edge (e.g., the mapping at +40 meV shown in Fig. 2(d)), the characteristics of extended electron wavefunctions gradually shows up with more homogeneous LDOS distribution. Statistically, the distribution of dI/dV values in the mapping (cluster regions removed) progressively evolves from a log-normal distribution (red dashed curves at -40 meV and -30 meV in Fig. 2(e)) to a Gaussian distribution (blue dashed curves at +30 meV and +40 meV in Fig. 2(f)). To eliminate the STM setup effect in the LDOS distributions, we further examine the mappings of (dI/dV)/(I/V) instead of dI/dV on the same regions (cluster regions removed), and the transition between log-normal and Gaussian distribution remains clear (see in Fig. S7 in Supporting Information). Such log-normal statistical distribution has been reported in disordered systems close to the critical point of Anderson transition [35-38]. The transition between extended states and localized states (Fig. 2 and Fig. S8, S9 in Supporting Information) indicates that the Fermi level is near the mobility edge of Anderson localization, and the corresponding crossover from Gaussian to log-normal distributions in the statistics of dI/dV values provides a glimpse into the multifractal nature of electron wavefunctions near SIT [39, 40]. It is worth noting that the log-normal distributions are indicative of localization effects. For even higher density of iron clusters, the enlarged height variations on the surface prevent us to obtain a decent LDOS mapping. In comparison, dI/dV mappings on monolayer FTS before depositing iron clusters show neither signatures of localization nor log-normal distributions in dI/dV mappings (see Fig. S10 in Supporting Information).

Further increasing the disorder level by iron clusters, stronger phase fluctuation and Anderson localization may result in a breakdown of global phase coherence of Cooper pairs, whereas local incoherent Cooper pairing still exist [10, 14, 18, 33]. Theoretically, localization-enhanced pairing may come into play, and consequently, strongly bound electron pairs form and survive deep into the insulating regime of SIT, leading to a large spectral gap of the "superconductivity-induced" insulator [5, 6]. In our study, after iron clusters are further deposited onto the film for the 3rd time, the cluster coverage is raised to 0.019 ~ 0.042 $nm^{-2}$ (see Fig. S5(c1)-(c5) in Supporting Information). In the regions of high cluster coverages ($\geq$ 0.036 $nm^{-2}$), we detect large asymmetric U-shaped spectral gaps as shown by Fig. 3(b)-(c). The topography corresponding to Fig. 3(b) is shown in Fig. 3(a). Such large spectral gap does not show coherence peaks and the gap size grows with the increasing cluster coverage (Fig. 3(b)-(c)), which can be indicative of the insulating regime of SIT. The U-shaped gap and the correlation between its size and the disorder level appear to be consistent with the theoretically predicted “superconductivity-induced” insulating gap originating from the enhanced Cooper pair correlations [6, 17].

Evolving from low to high disorder levels, Figure 4 summarizes three types of spectra observed with the most representative characteristics, namely the spectra with superconducting gaps, V-shaped gaps, and large U-shaped gaps (corresponding to the cases in Figs. 1(b), 1(g) and 3(b-c) respectively). The spectral evolution is indicative of a quantum phase transition, which is supported by our transport measurements (Fig. S12). We note that the large U-shaped gap at the high disorder level is consistent with numerical studies that model the interplay between superconductivity and localization [6, 7]. The numerical studies suggest that the disorder effect confines electron pairs to a smaller localization volume, consequently inducing a large U-shaped gap. The V-shaped gap at the intermediate disorder level seems to be different from the predicted hard gap at the SIT quantum critical point in Ref. [[6, 7]]. This discrepancy might come from finite temperature broadening or electron-electron interaction effects ignored in the model calculations. Besides, the numerical model does not account for the substantial increase in the spectral weight at positive energies as temperature rises (Fig. 3). At high disorder levels, the spectra with large U-shaped gaps prevail (Fig. 4(h)), which are absent at low disorder levels. Spectra with small U-shaped gap features can also be observed occasionally, mixing with the large U-shaped gaps (Fig. S13 in Supporting Information). Our spectroscopic results above are consistent with the scenario of superconducting puddles [7, 8], indicative of a disorder-induced quantum phase transition [15, 19] (Fig. S14 in Supporting Information).

In conclusion, we have performed systematic spectroscopic study of the disorder-induced breakdown of superconductivity in high-$T_C$ superconducting monolayer FTS, where the disorder level is controlled through *in situ* depositing iron clusters by MBE. With increasing coverage of iron clusters, an evolution in the spectral features from superconducting gaps to V-shaped gaps, and finally to large U-shaped gaps is observed. The profound impact of disorder is shown by the transition in dI/dV mappings from extended states to localized states, indicating that the mobility edge of Anderson localization is near the Fermi level and interact with superconductivity. The large U-shaped gap growing with the increasing amount of disorder is detected on monolayer FTS, which is a signature of the insulating regime of SIT that has never been spectroscopically explored before. The formation of such large gap in the strong disorder regime should be related to the theoretically predicted enhanced pairing correlation within the localization length, reflecting an intriguing interplay between superconductivity and localization. We expect the results presented here will stimulate further studies into the role of disorder in quantum phase transitions of unconventional superconductors.

## *Acknowledgments*

We thank Haiwen Liu for helpful discussion. G. H. and Z. W. contributed equally to this letter. This work was supported by the National Natural Science Foundation of China (Grant No. 12488201), the National Key Research and Development Program of China (No. 2023YFA1406500, No. 2022YFA1403103), the Shanghai Baiyulan Talent Plan Pujiang Project (Grant No. 24PJA081), the Quantum Science and Technology-National Science and Technology Major Project (Grant No. 2021ZD0302403), and the National Natural Science Foundation of China (No. 12174442).

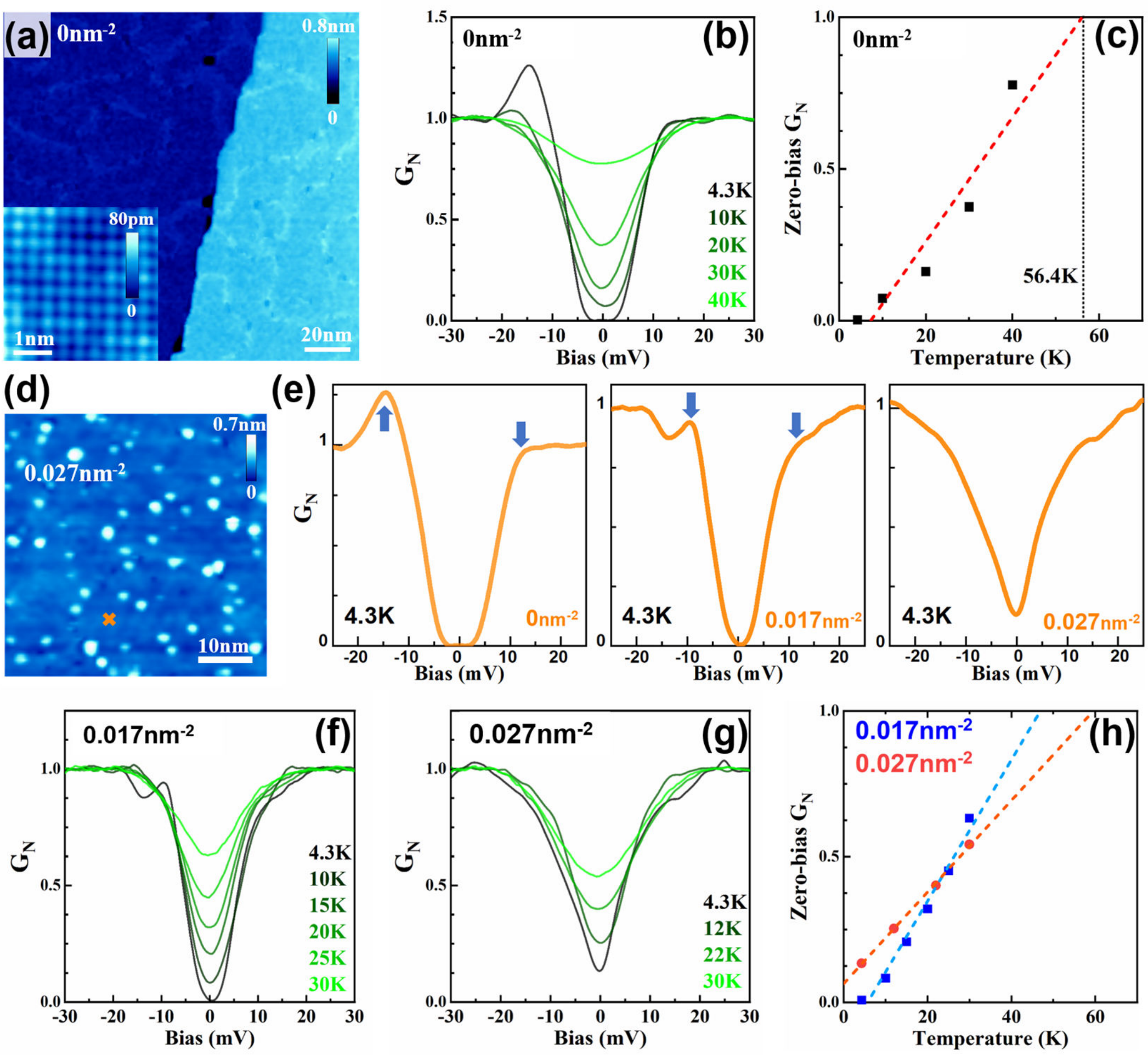

FIG. 1 (a) STM topography of the pristine monolayer FTS grown on STO terraces (higher terrace in brighter color), with the atomically resolved image in the inset. (b) Normalized spectra $G_N$ of the pristine monolayer FTS in (a) at varying temperatures. (c) $T^*$ obtained by extrapolating the zero-bias $G_N$ (black dots) of each spectrum in (b). (d) STM topography of the monolayer FTS with iron clusters deposited as bright spots, where tunneling spectra are measured at the cross-marked position. (e) Normalized spectra $G_N$ at 4.3 K for different surface cluster coverages (number of clusters per area), with the coherence peaks denoted by blue arrows. Spectra for 0.027 $nm^{-2}$ coverage are measured at the position marked by a cross in (d). (f)-(g) Normalized spectra of the monolayer FTS with two different surface cluster coverages at varying temperatures. (h) Extrapolating the zero-bias conductance (blue/red dots) of each spectrum in (f) or (g) respectively to obtain $T^*$.

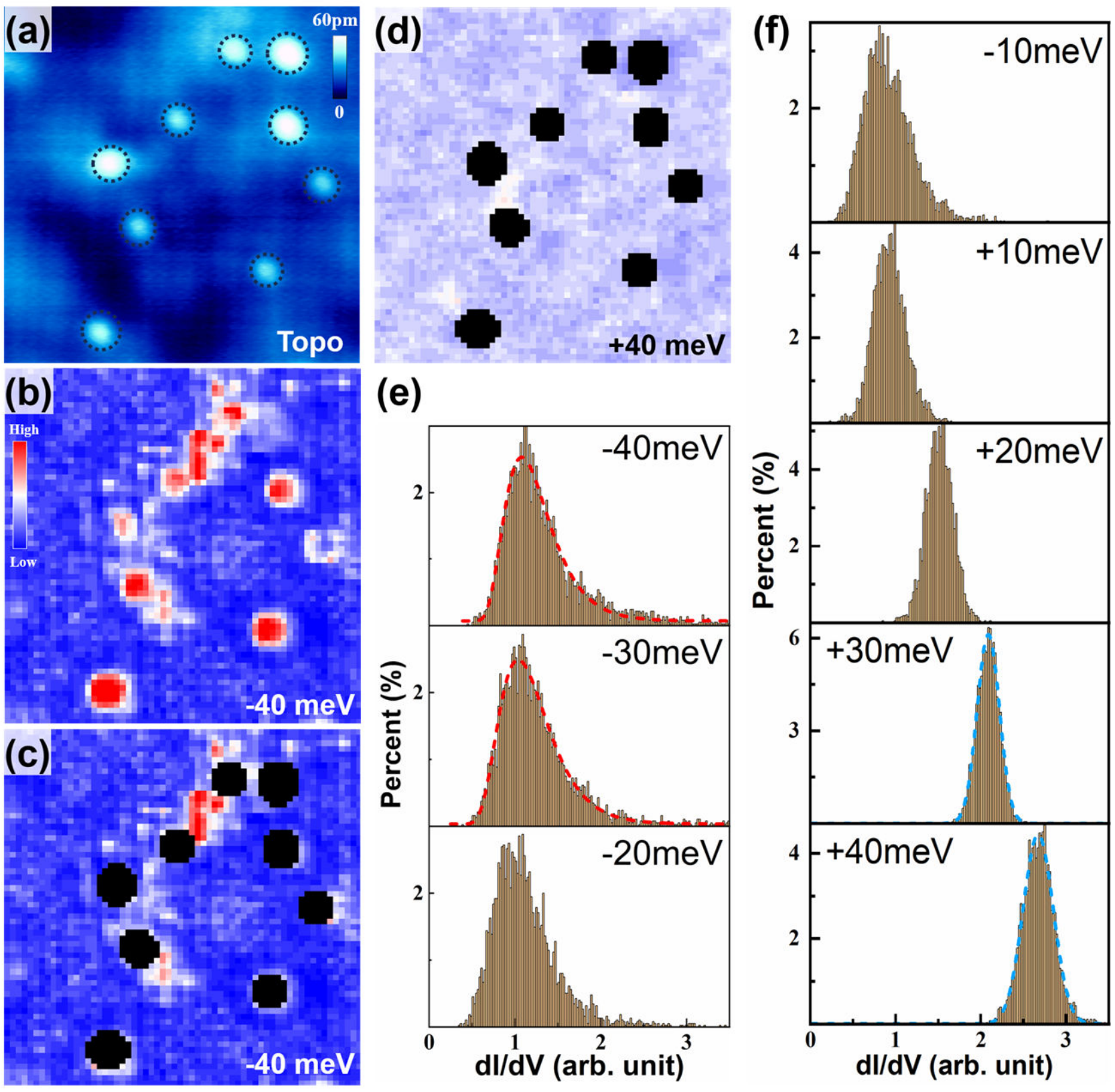

FIG. 2 (a) STM topography of an area on monolayer FTS with an accumulation of iron clusters, whose dI/dV mappings are shown in (b-d). After excluding the dI/dV values at the positions of iron clusters (dashed circles in (a)), the dI/dV mapping at -40 meV in (b) is shown in (c) with excluded regions in black color. (d) dI/dV mapping at +40 meV with the same cluster regions excluded. (e-f) Statistical distributions of the dI/dV values in (c-d) and other energies, with cluster regions excluded. The Gaussian and log-normal fits are denoted by blue and red dashed curves respectively. dI/dV mappings at other energies are shown in Fig. S11 in Supporting Information.

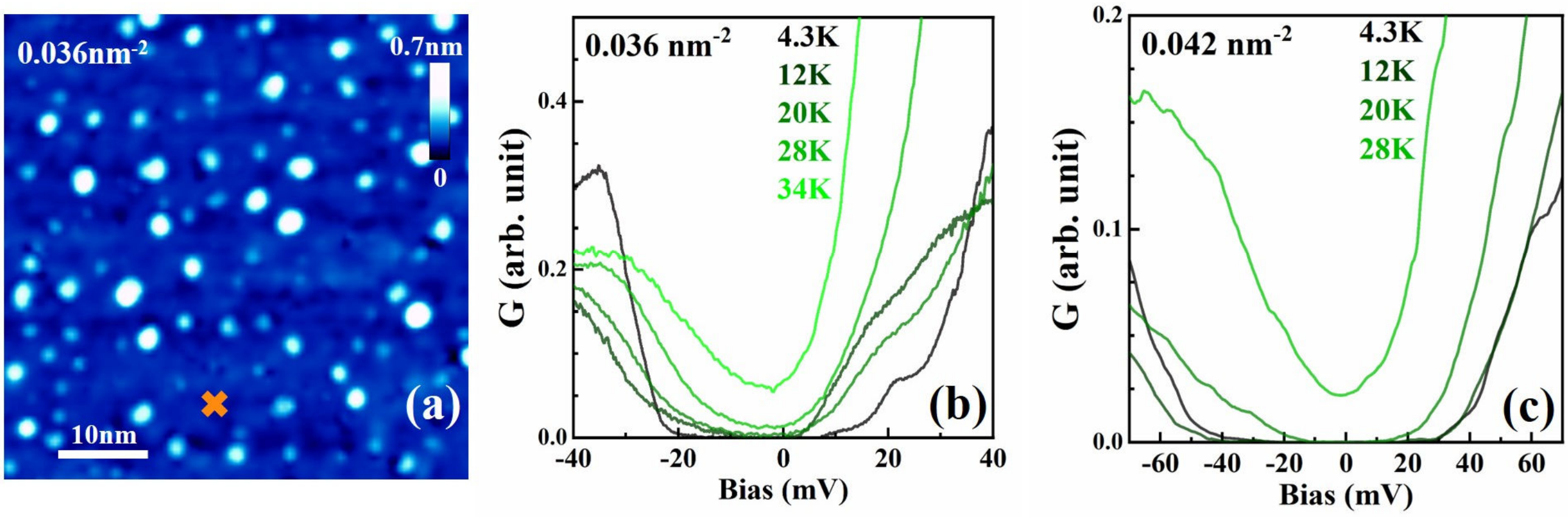

FIG. 3 STM topography and STS spectra in the monolayer FTS with high cluster coverage. (a) STM topography of the region on monolayer FTS at a relatively high disorder level with the iron cluster coverage of 0.036 $nm^{-2}$. (b) Raw spectra of the monolayer FTS at varying temperatures, measured at the position marked by the cross in (a). (c) Spectra for regions on the monolayer FTS of higher disorder levels with the iron cluster coverage of 0.042 $nm^{-2}$. The STS set-up of the spectra in these figures is V = 100 mV, I = 2.0 nA. The observed large gaps indicate the insulating states.

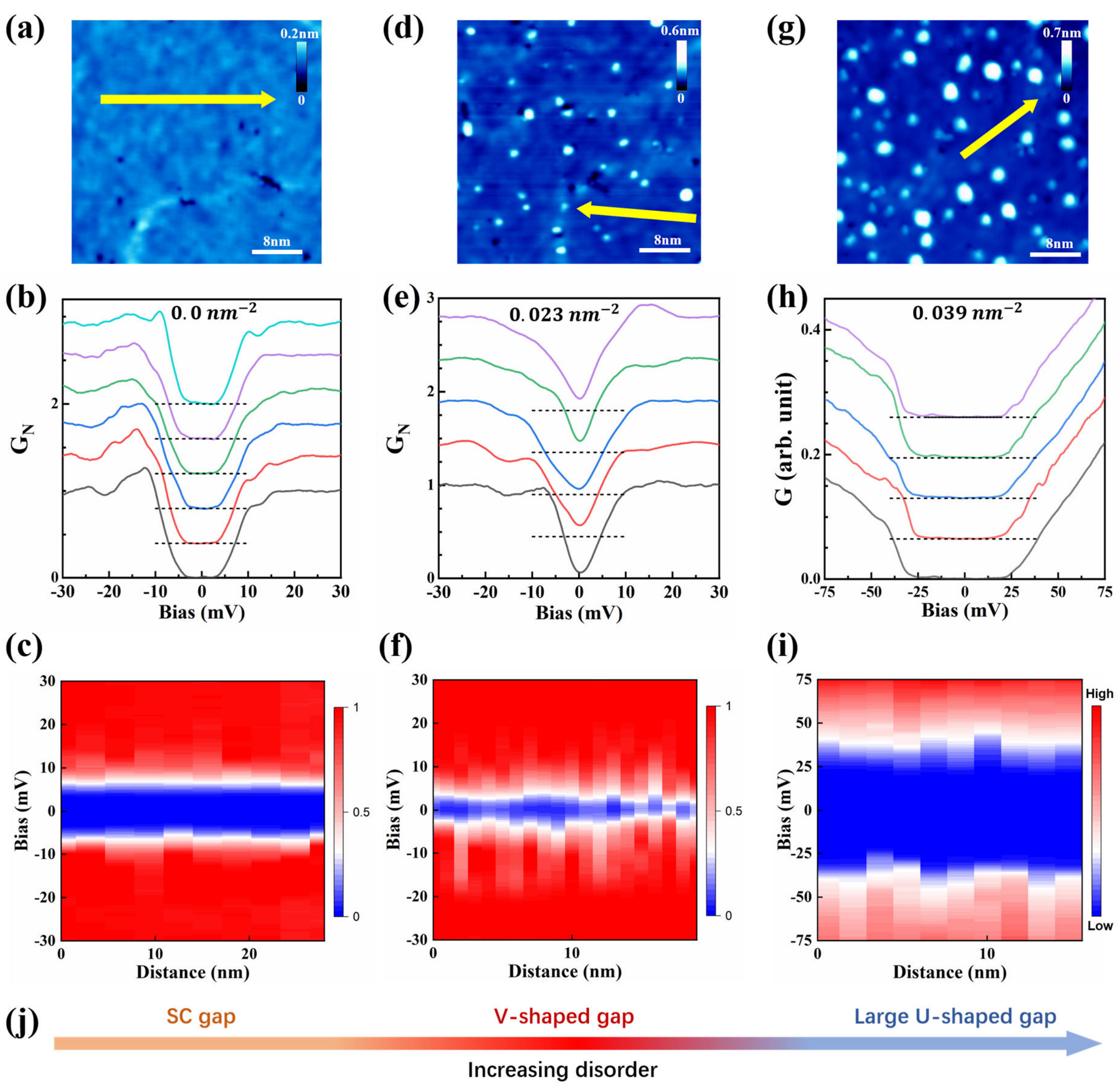

FIG. 4 Spatial distribution of spectral gaps at different disorder levels. (a) A typical region of monolayer FTS before the deposition of iron clusters, where tunneling spectra are measured along the linecut denoted by the yellow arrow. (b) Several spectra measured along the linecut in (a). (c) Color-scale representation of all spectra along the linecut in (a), where blue colors represent low values of dI/dV (energy gap) and red colors represent high dI/dV. (a)-(c) show the spatial distribution of superconducting spectra. Similarly, (d)-(f) and (g)-(i) show those with V-shaped gaps at moderate disorder levels and U-shaped gaps at high disorder levels, respectively. In figures (b), (e) and (h), the iron cluster coverage is labelled, and the spectra are vertically offset for clarity, with the dashed lines indicating the zero-conductance positions. (j) Evolution from superconducting gaps with coherence peaks to V-shaped gaps and to large U-shaped gaps with increasing disorder, corresponding to the disorder-induced breakdown of superconductivity.